LA-UR-01-5346

*Approved for public release; distribution is unlimited.*

*Title:*

# Experimental and Theoretical Study of the Residual Product Nuclide Yields in 100–2600 MeV Proton-Irradiated Thin Targets

*Author(s):*

Yury E. TITARENKO, Vyacheslav F. BATYAEV, Evgeny I. KARPIKHIN, Valery M. ZHIVUN, Aleksander B. KOLDOBSKY, Ruslan D. MULAMBETOV, Svetlana V. KVASOVA, Dmitry V. FISCHENKO, Vladilen S. BARASHENKOV, Stepan G. MASHNIK, Richard E. PRAEL, Arnold J. SIERK, Hideshi YASUDA, Masaki SAITO

*Submitted to:*

# Experimental and Theoretical Study of the Residual Product Nuclide Yields in 100–2600 MeV Proton-Irradiated Thin Targets

Yury E. TITARENKO[1,*], Vyacheslav F. BATYAEV[1], Evgeny I. KARPIKHIN[1], Valery M. ZHIVUN[1], Aleksander B. KOLDOBSKY[1], Ruslan D. MULAMBETOV[1], Svetlana V. KVASOVA[1], Dmitry V. FISCHENKO[1], Vladilen S. BARASHENKOV[2], Stepan G. MASHNIK[3], Richard E. PRAEL[3], Arnold J. SIERK[3], Hideshi YASUDA[4], Masaki SAITO[5]

[1] *Institute for Theoretical and Experimental Physics, B.Cheremushkinskaya 25, 117259 Moscow, Russia*

[2] *Joint Institute for Nuclear Research, 141980, Dubna, Moscow reg., Russia*

[3] *Los Alamos National Laboratory, Los Alamos, NM 87545, USA*

[4] *Japan Atomic Energy Research Institute, Tokai-mura, Ibaraki 319-1195, Japan*

[5] *Tokyo Institute of Technology 2-12-1, O-okayama, Meguro-ku, Tokyo 152, Japan*

The work is aimed at experimental determining and computer simulating the independent and cumulative yields of residual product nuclei in the target and structure materials of the transmutation facilities driven by high-current accelerators. The ITEP U-10 accelerator was used in 48 experiments to obtain more than 4000 values of the yields of radioactive residual product nuclei in 0.1-2.6 GeV proton-irradiated thin $^{182,183,184,186}$W , $^{nat}$W, $^{56}$Fe, $^{58}$Ni, $^{93}$Nb, $^{232}$Th, $^{nat}$U, $^{99}$Tc, $^{59}$Co $^{63,65}$Cu, $^{nat}$Hg, $^{208}$Pb, and $^{27}$Al targets. The results of verifying the LAHET, CEM95, CEM2k, CASCADE, CASCADE/INPE, YIELDX, HETC, INUCL, and other simulation codes are presented.

## I. Introduction

In recent years, the residual product yield data have been widely adopted in the feasibility analyses of accelerator-driven systems (ADS) applicable, for instance, to nuclear waste transmutation[1)] . This is related primarily to the information on the applicability scope of the various simulation codes used to calculate high-energy interactions in the ADS structure elements with a view to more reliable calculations of the ADS nuclear parameters.

Table 1 lists the proton energies and the target materials studied in the work.

**Table 1** Target materials and proton energies

| Target | Proton energy (GeV) | | | | | | |
|---|---|---|---|---|---|---|---|
| | 0.1 | 0.2 | 0.8 | 1.0 | 1.2 | 1.6 | 2.6 |
| $^{182}$W | | √ | √ | | | √ | |
| $^{183}$W | | √ | √ | | | √ | |
| $^{184}$W | | √ | √ | | | √ | |
| $^{186}$W | | √ | √ | | | √ | |
| $^{232}$Th | √ | √ | √ | | √ | √ | |
| $^{nat}$U | √ | √ | √ | | √ | √ | |
| $^{99}$Tc | √ | √ | √ | | √ | √ | |
| $^{59}$Co | | √ | | | √ | √ | √ |
| $^{63}$Cu | | √ | | | √ | √ | √ |
| $^{65}$Cu | | √ | | | √ | √ | √ |
| $^{nat}$Hg | √ | √ | √ | | | | √ |
| $^{56}$Fe | | | | | | | √ |
| $^{58}$Ni | | | | | | | √ |
| $^{93}$Nb | | | | | | | √ |
| $^{nat}$W | | | | | | | √ |
| $^{208}$Pb | | | | √ | | | |

* Corresponding author, Tel. +7-095-123-6383, Fax. +7- 095-127 -0543, E-mail: Yury.Titarenko@itep.ru

## II. Experiment

The samples of 10.5 mm diameter were irradiated by the external proton beam from the ITEP U-10 synchrotron[2)] . The nuclide yields were determined by the direct $\gamma$-spectrometry method. The $\gamma$-spectrometer resolution is 1.8 keV in the 1332 keV $\gamma$-line. The $\gamma$-spectra were processed by the GENIE2000 code. The $\gamma$-lines were identified, and the cross sections calculated, by the ITEP-developed SIGMA code using the PC-NUDAT database. The proton fluence was monitored by the $^{27}$Al(p,x)$^{22}$Na reaction.

A more detailed description of the experimental techniques can be found in[2)] and[3)] .

## III. Basic definitions and computational relations

The formalism of representing the reaction product yields (cross sections) in high-energy proton-irradiated thin targets is described in sufficient detail in[2)] . In terms of the formalism, the variations in the concentration of any two chain nuclides produced in an irradiated target ($N_1 \xrightarrow{\lambda_1} N_2 \xrightarrow{\lambda_2}$) may be presented to be a set of differential equations that describe the production and decays of the nuclides. By introducing a formal representation of the time functions of the type $F_i = \left(1 - e^{-\lambda_i \tau}\right) \dfrac{1 - e^{-\lambda_i KT}}{1 - e^{-\lambda_i T}}$ , (i=1, 2, Na; $\tau$ is the duration of accelerated proton pulse; $T$ is the pulse repetition period; $k$ is the number of pulses within the irradiation period), which characterize the nuclide decays within the irradiation time, and by expressing (similar to the relative measurements) the proton fluence via monitor reaction cross section $\sigma_{st}$, we can present the unknowns as

$$\sigma_1^{cum} = \frac{A_0}{\eta_1 \varepsilon_1 F_1 N_{Na}} \frac{N_{Al}}{N_T} \frac{F_{Na}}{\lambda_{Na}} \sigma_{st}, \qquad (1)$$

$$\sigma_1^{cum} = \frac{A_1}{\nu_1 \eta_2 \varepsilon_2 F_1 N_{Na}} \frac{N_{Al}}{N_T} \frac{\lambda_2 - \lambda_1}{\lambda_2} \frac{F_{Na}}{\lambda_{Na}} \sigma_{st}, \qquad (2)$$

$$\sigma_2^{ind} = \left(\frac{A_2}{F_2} + \frac{A_1}{F_1}\frac{\lambda_1}{\lambda_2}\right)\frac{1}{\eta_2\varepsilon_2 N_{Na}}\frac{N_{Al}}{N_T}\frac{F_{Na}}{\lambda_{Na}}\,\sigma_{st}, \quad (3)$$

$$\sigma_2^{cum} = \sigma_2^{ind} + \nu_1\sigma_1^{cum} = \left(\frac{A_1}{F_1} + \frac{A_2}{F_2}\right)\frac{1}{\eta_2\varepsilon_2 N_{Na}}\frac{N_{Al}}{N_T}\frac{F_{Na}}{\lambda_{Na}}\,\sigma_{st}\,, \quad (4)$$

where $\sigma_1^{cum}$ is the cumulative cross section of the first nuclide; $\sigma_2^{ind}$ and $\sigma_2^{cum}$ are the independent and cumulative cross sections of the second nuclide; $N_{Al}$ and $N_T$ are the numbers of nuclei in the monitor (standard) and in experimental sample, respectively; $\eta_1$ and $\eta_2$ are the $\gamma$-line yields; $\varepsilon_1$ and $\varepsilon_2$ are the spectrometer efficiencies at energies $E_{\gamma_1}$ and $E_{\gamma_2}$; $\nu_1$ is the branching ratio of the first nuclide; $\lambda_1$, $\lambda_2$, $\lambda_{Na}$ are, respectively, the decay constants of the first and second nuclides and of the monitor product ($^{22}$Na). The factors $A_0$, $A_1$, and $A_2$ are calculated through fitting the measured counting rates in the total absorption peaks, which correspond to energies $E_{\gamma_1}$ (the first nuclide) and $E_{\gamma_2}$ (the second nuclide), by exponential functions.

A more detailed description of the computational relations can be found in[3)] and[4)] .

## IV. Experimental results

More than four thousand yields of reaction products in the targets listed in Table 1 have been determined and presented in our Report[3)] . The data will be sent to the EXFOR experimental database. Table 2 shows the total number of the measured reaction product yields of different types in each of our measurement runs.

## V. Simulation of experimental data

The following thirteen codes were used to simulate the experimental data, thus permitting the predictive power of the codes to be estimated. CEM95,[5)] CEM2k,[6)] CASCADE,[7)] INUCL,[8)] LAHET,[9)] YIELDX,[10)] CASCADE/INPE,[11)] CASCADO/IPPE,[12)] GNASH,[13)] ALICE with the Fermi/Kataria distribution of the nuclear level densities,[14)] QMD,[15)] NUCLEUS,[16)] and ALICE-IPPE[17)] .

The procedure of calculating the cumulative yields and comparing between the experimental and simulation data is described in detail in.[2–4)]

As an example, Fig. 1 shows the results of a detailed comparison between simulated and experimental independent and cumulative products in $^{208}$Pb irradiated with 1GeV protons.

## VI. Summary on the agreement between the experimental and simulated product nuclide yields

The comparison was made for two groups of nuclei, namely, the groups with a significant fission mode (conditionally "heavy" nuclei, $^{182,183,184,186,nat}$W, $^{nat}$Hg, $^{208}$Pb, $^{232}$Th, and $^{nat}$U) and without any fission mode (conditionally "light" nuclei, $^{56}$Fe, $^{58}$Ni, $^{59}$Co, $^{63,65}$Cu, $^{93}$Nb, and $^{99}$Tc). Figure 2 reflects the information on the predictive power of the codes (mean squared deviation factor is presented).

In the case of light nuclei, where almost all product nuclides are formed by spallation, the predictive power of most of the Monte-Carlo codes is characterized by an mean squared deviation factor of at least 2.0, with the agreement being somewhat worse at low energies. The YIELDX semi-phenomenological code gives the best result when predicting the reaction product yields in light nuclei and sometimes approaches the required 30% accuracy[18)] .

**Table 2** Number of measured reaction product yields of different types in each experiment

| Experiment | | Yield type | | | | Total |
|---|---|---|---|---|---|---|
| Target | $E_p$ [GeV] | i | c | i* | i** | |
| $^{182}$W | 0.2 | 3 | 25 | 1 | 3 | 32 |
| $^{182}$W | 0.8 | 5 | 58 | 1 | 6 | 70 |
| $^{182}$W | 1.6 | 10 | 87 | 6 | 6 | 109 |
| $^{183}$W | 0.2 | 4 | 26 | 1 | 4 | 35 |
| $^{183}$W | 0.8 | 6 | 61 | 2 | 7 | 76 |
| $^{183}$W | 1.6 | 12 | 87 | 6 | 6 | 111 |
| $^{184}$W | 0.2 | 4 | 26 | 1 | 5 | 36 |
| $^{184}$W | 0.8 | 7 | 61 | 2 | 7 | 77 |
| $^{184}$W | 1.6 | 12 | 88 | 7 | 7 | 114 |
| $^{186}$W | 0.2 | 4 | 26 | 1 | 5 | 36 |
| $^{186}$W | 0.8 | 4 | 53 | 1 | 4 | 62 |
| $^{186}$W | 1.6 | 13 | 90 | 8 | 8 | 119 |
| $^{nat}$W | 2.6 | 10 | 104 | 9 | 6 | 129 |
| $^{232}$Th | 0.1 | 10 | 60 | 9 | 8 | 87 |
| $^{232}$Th | 0.2 | 16 | 84 | 18 | 10 | 128 |
| $^{232}$Th | 0.8 | 15 | 89 | 15 | 11 | 130 |
| $^{232}$Th | 1.2 | 22 | 153 | 19 | 20 | 214 |
| $^{232}$Th | 1.6 | 22 | 156 | 18 | 16 | 212 |
| $^{nat}$U | 0.1 | 12 | 77 | 9 | 10 | 108 |
| $^{nat}$U | 0.2 | 15 | 80 | 15 | 13 | 123 |
| $^{nat}$U | 0.8 | 21 | 137 | 17 | 20 | 195 |
| $^{nat}$U | 1.2 | 22 | 161 | 22 | 21 | 226 |
| $^{nat}$U | 1.6 | 23 | 166 | 22 | 20 | 231 |
| $^{99}$Tc | 0.1 | 4 | 9 | 3 | 2 | 18 |
| $^{99}$Tc | 0.2 | 4 | 21 | 9 | 5 | 39 |
| $^{99}$Tc | 0.8 | 10 | 43 | 11 | 8 | 72 |
| $^{99}$Tc | 1.2 | 8 | 41 | 12 | 6 | 67 |
| $^{99}$Tc | 1.6 | 10 | 47 | 11 | 10 | 78 |
| $^{59}$Co | 0.2 | 6 | 17 | 3 | 3 | 29 |
| $^{59}$Co | 1.2 | 7 | 26 | 4 | 4 | 41 |
| $^{59}$Co | 1.6 | 7 | 26 | 4 | 4 | 41 |
| $^{59}$Co | 2.6 | 7 | 26 | 4 | 4 | 41 |
| $^{63}$Cu | 0.2 | 9 | 13 | 3 | 4 | 29 |
| $^{63}$Cu | 1.2 | 10 | 29 | 4 | 4 | 47 |
| $^{63}$Cu | 1.6 | 11 | 23 | 4 | 4 | 42 |
| $^{63}$Cu | 2.6 | 11 | 23 | 4 | 4 | 42 |
| $^{65}$Cu | 0.2 | 8 | 14 | 4 | 3 | 29 |
| $^{65}$Cu | 1.2 | 13 | 31 | 5 | 5 | 54 |
| $^{65}$Cu | 1.6 | 10 | 27 | 5 | 5 | 47 |
| $^{65}$Cu | 2.6 | 10 | 28 | 5 | 5 | 48 |
| $^{nat}$Hg | 0.1 | 4 | 18 | 10 | 12 | 44 |
| $^{nat}$Hg | 0.2 | 6 | 33 | 12 | 14 | 65 |
| $^{nat}$Hg | 0.8 | 9 | 68 | 12 | 14 | 103 |
| $^{nat}$Hg | 1.6 | 8 | 103 | 16 | 14 | 141 |
| $^{56}$Fe | 2.6 | 5 | 25 | 3 | 3 | 36 |
| $^{58}$Ni | 2.6 | 9 | 22 | 3 | 4 | 38 |
| $^{93}$Nb | 2.6 | 6 | 58 | 12 | 8 | 85 |
| $^{208}$Pb | 1.0 | 8 | 76 | 15 | 15 | 114 |
| | **Total** | **472** | **2802** | **388** | **387** | **4050** |

**i** – independent yields of ground states, **c** – cumulative yields,
**i*** = $\mathbf{i}_{\Sigma m_j}$ – independent yields of metastable states,
**i**** = $\mathbf{i}_{\Sigma m_j + g}$ – summed-up independent yields of metastable and ground states.

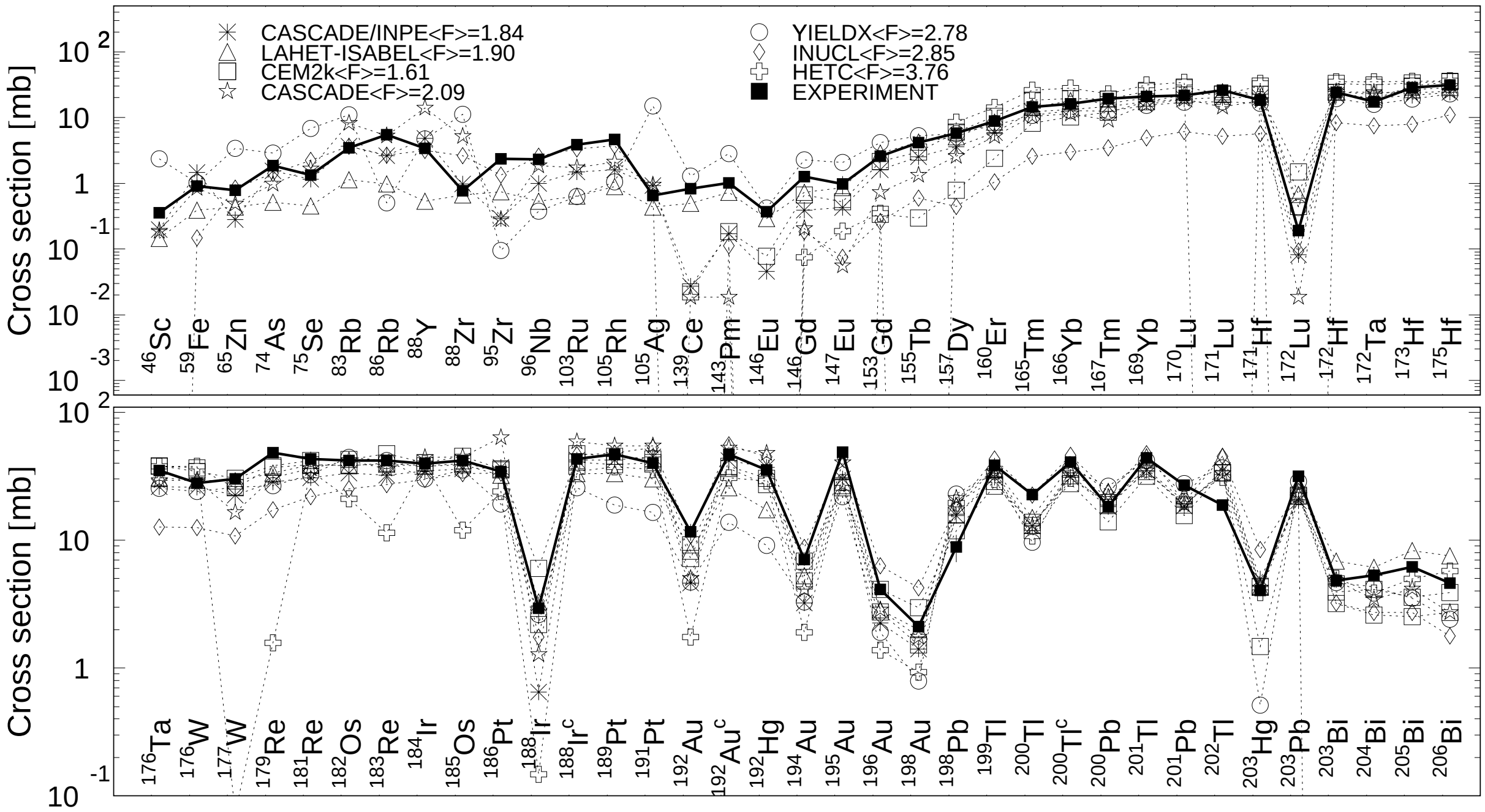

**Fig. 1** Detailed comparison between experimental and simulated yields of radioactive reaction products in 1GeV proton-irradiated $^{208}$Pb. The cumulative yields are labeled with a "c" when the respective independent yields are also shown. The mean simulation-to-experimental data ratios <F> are shown in the legend for each of the codes.

In the case of heavy nuclei, the physics of proton-nucleus interactions gets complicated due to the fission channel which is not even included in some of the tested codes (CEM95, CEM2k, HETC). Therefore, the mean squared deviation factor is very high (commonly, at least 3.0 and sometimes about an order of magnitude) for the fission products. From this it follows that, although the spallation products are described by the present-day codes somewhat better for heavy nuclei that for light nuclei (the mean squared deviation factor is bellow 2.0), the general agreement is about the same as in the case of light nuclei (the mean squared deviation factor is about 2.0 and higher). Our study shows that further development of reliable fission models is a priority task in updating all the simulation codes.

It should be also noted that, in the case of high-energy ($E_p$ >1 GeV) projectile protons, most of the tested codes fail to satisfactorily describe the production of the nuclides whose nucleon compositions are close to the primary nuclei. This indicates that the the physical models used to describe $(p, xpyn)$–type processes are imperfect when $x + y \leq 3$.

## VII. Conclusion

The presented results show that almost all of the above-verified codes are applicable during the stage of conceptual feasibility study and development work, but are not yet reliable enough to solve the applied problems that arise when designing and operating the ADS facilities. At the same time, the yields of numerous secondary products have to be known to within a very high accuracy for many reasons (large cross sections for neutron capture, a high radiotoxicity, chemical poisoning of structure elements, gas evolution, etc.). So, the codes have to be much improved to become a reliable tool for calculating the ADS parameters. The experience gained in the present researches has shown that the experiment-simulation differences are expedient to study in detail when the projectile proton energy range is broken into several intervals. This approach will be realized for the Pb and Bi targets in the ISTC Project #2002

## Acknowledgment

The authors are indebted to Dr. F. E. Chukreev for his helpful comments on the nuclear decay-chain data. Also we thank Drs. S. Chiba, H. Takada, M. Blann, M. Chadwick, T. Gabriel, Yu. N. Shubin, and A. V. Ignatyuk for the theoretical calculations made via their codes.

The work has been carried out under the ISTC Project #839[3)] supported by the European Community, Japan (JAERI), and Norway. In part, the work has been supported by the U.S. Department of Energy.

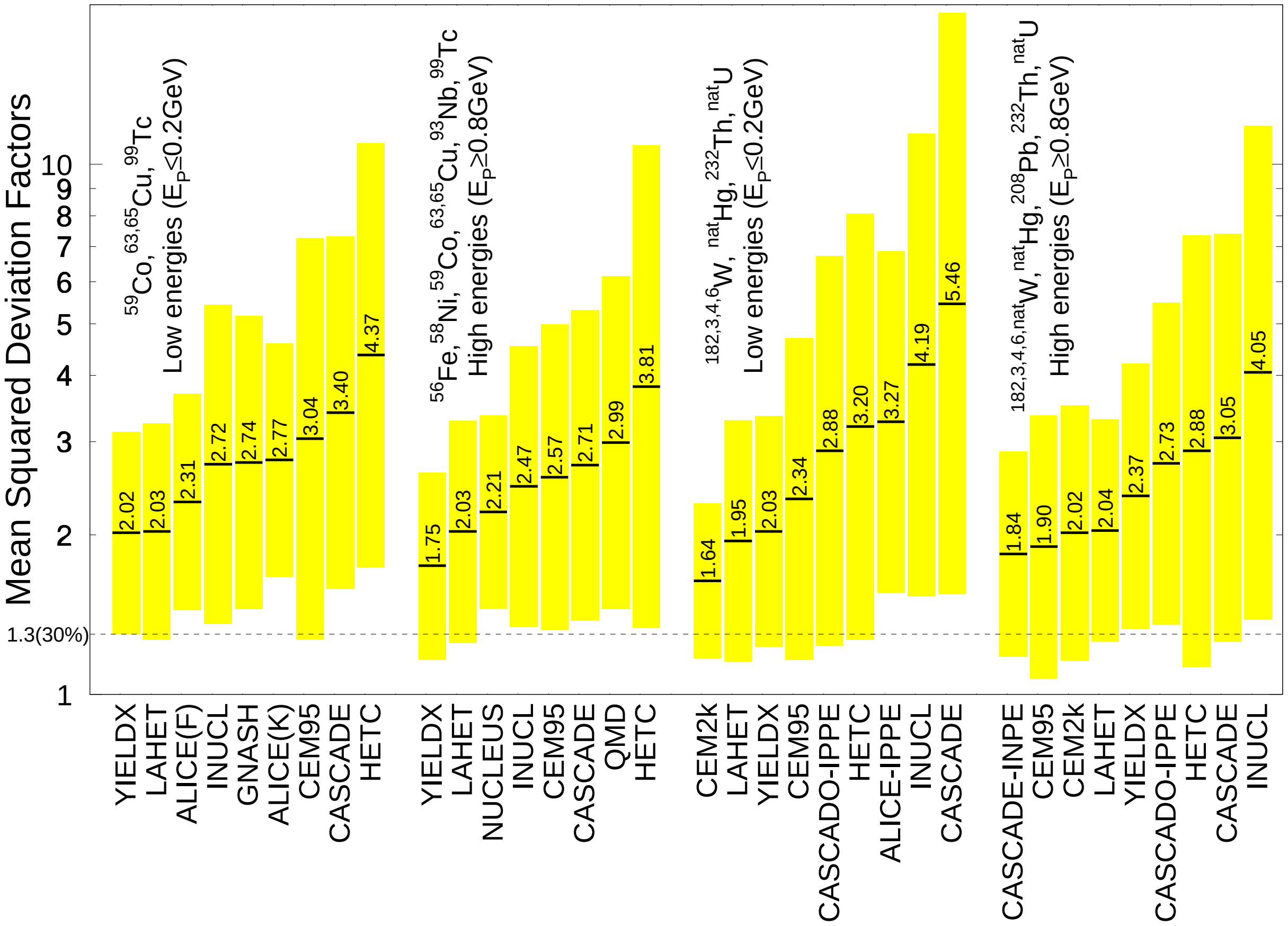

**Fig. 2** The mean squared deviation factor for the unified comparison.